\documentclass[a4paper,12pt,reqno]{article}
\usepackage{amsmath,amsfonts,amssymb,amsthm,dsfont}

\usepackage{hyperref}
\allowdisplaybreaks
\numberwithin{equation}{section}

\newcommand{\Om}{\Omega}
\newcommand{\om}{\omega}
\newcommand{\half}{\tfrac{1}{2}}
\newcommand{\ihalf}{\tfrac{\mathrm{i}}{2}}

\newcommand{\Ii}{\mathrm{i}}

\newcommand{\6}{\partial}
\newcommand{\7}{\hat}

\newcommand{\ua}{{\underline{\alpha}}}
\newcommand{\ub}{{\underline{\beta}}}
\newcommand{\ug}{{\underline{\gamma}}}
\newcommand{\ud}{{\underline{\delta}}}
\newcommand{\ul}[1]{{\underline{#1}}}

\newcommand{\oa}{{\overline{\alpha}}}

\newcommand{\com}[2]{[\,#1\, ,\,#2\,]}	
\newcommand{\acom}[2]{\{#1\, ,\,#2\}}	

\newcommand{\gam}{\Gamma}	
\newcommand{\GAM}{\hat{\Gamma}}	
\newcommand{\CC}{C}	
\newcommand{\IC}{C^{-1}}	

\newcommand{\Dim}{D} 
\newcommand{\Ns}{N} 
\newcommand{\Omg}{\Omega_\mathrm{gh}} 
\newcommand{\fb}{s_\mathrm{gh}} 
\newcommand{\Hg}{H_\mathrm{gh}} 
\newcommand{\hHg}{\7H_\mathrm{gh}} 
\newcommand{\cdeg}{$c$-degree} 

\newcommand{\xx}[1]{\xi^{\ul{#1}}}

\newcommand{\tc}[1]{\tilde{c}^{\,#1}}

\newcommand{\LRA}{\Leftrightarrow}

\newcommand{\so}[1]{$\mathfrak{so}(#1)$}
\newcommand{\pap}{\cite{Brandt:2009xv}}

\newtheorem{prop}{Proposition}[section]
\newtheorem{lemma}[prop]{Lemma}

\begin{document}

\begin{flushright}
ITP--UH--06/10
\end{flushright}

\begin{center}
 {\large\bfseries Supersymmetry algebra cohomology II:\\[6pt] Primitive elements in 2 and 3 dimensions}
 \\[5mm]
 Friedemann Brandt \\[2mm]
 \textit{Institut f\"ur Theoretische Physik, Leibniz Universit\"at Hannover, Appelstra\ss e 2, D-30167 Hannover, Germany}
\end{center}

\begin{abstract}
The primitive elements of the supersymmetry algebra cohomology as defined in a companion paper are computed exhaustively for standard supersymmetry algebras in dimensions $\Dim=2$ and $\Dim=3$, for all signatures $(t,\Dim-t)$ and all numbers $\Ns$ of sets of supersymmetries.
\end{abstract}

\tableofcontents

\section{Introduction}

This paper relates to supersymmetry algebra cohomology as defined in \pap, for supersymmetry algebras in dimensions $\Dim=2$ and $\Dim=3$ of translational generators $P_a$ ($a=1,\dots,\Dim)$ and supersymmetry generators $Q^i_\ua$ of the form
 \begin{align}
  \com{P_a}{P_b}=0,\quad \com{P_a}{Q^i_\ua}=0,\quad \acom{Q^i_\ua}{Q^j_\ub}=-\Ii\,\delta^{ij}\,(\gam^a \IC)_{\ua\ub}P_a
	\label{i2-1} 
 \end{align}
where $\delta^{ij}$ denotes the Kronecker delta for $\Ns$ sets of supersymmetries,\footnote{The index $i=1,\dots,\Ns$  numbers sets of supersymmetries. In the case of Majorana-Weyl supersymmetries we use $i=1_+,\dots,\Ns_+,1_-,\dots,\Ns_-$ with $\Ns=\Ns_++\Ns_-$ where the subscripts $+$ and $-$ indicate the chirality of the supersymmetries respectively.}
  \begin{align}
  \delta^{ij}=
  \left\{\begin{array}{rl}1 & \mbox{if\ $i=j$}\\
  0& \mbox{if\ $i\neq j$,}\end{array}\right.
	\label{i2-2} 
 \end{align}
and $\CC$ is a charge conjugation matrix fulfilling in all cases under study
 \begin{align}
  \forall a:\quad\CC\,\gam^a\IC=-\gam^{a\top}
	\label{i2-4} 
 \end{align}
and
 \begin{align}
  \CC^{\top}=-\CC.
	\label{i2-5} 
 \end{align}

The object of this paper is the determination of the primitive elements of the supersymmetry algebra cohomology for supersymmetry algebras \eqref{i2-1} in $\Dim=2$ and $\Dim=3$ dimensions, for all numbers $\Ns$ of sets of supersymmetries and all signatures $(t,\Dim-t)$ of the Clifford algebra of the gamma matrices $\gam^a$.  
According to the definition given in \pap, these primitive elements are the representatives of the cohomology $\Hg(\fb)$ of the coboundary operator
 \begin{align}
  \fb=\ihalf\, \delta^{ij}\,(\gam^a \IC)_{\ua\ub}\,\xi^\ua_i\xi^\ub_j\,\frac{\6}{\6 c^a}
	\label{i2-6} 
 \end{align}
in the space $\Omg$ of polynomials in translation ghosts $c^a$ and supersymmetry ghosts $\xi^\ua_i$ corresponding to the translational generators $P_a$ and the supersymmetry generators $Q^i_\ua$ respectively,
 \begin{align}
  &\Omg=\Big\{\sum_{p=0}^\Dim\sum_{n=0}^r c^{a_1}\dots c^{a_p}\xi^{\ua_1}_{i_1}\dots\xi^{\ua_n}_{i_n}
  a^{i_1\dots i_n}_{\ua_1\dots\ua_n a_1\dots a_p}\,|\,
  a^{i_1\dots i_n}_{\ua_1\dots\ua_n a_1\dots a_p}\in\mathbb{C},\ r=0,1,2,\ldots\Big\}.
	\label{i2-7} 
 \end{align}

Depending on the dimension $\Dim$ and on the signature $(t,\Dim-t)$ the supersymmetry generators and the supersymmetry ghosts are Majorana or symplectic Majorana spinors defined 
according to sections 2 and 4 of \pap\ by means of a matrix $B$ and, in the case of symplectic Majorana spinors, a matrix $\Om$: 
 \begin{align}
  \mbox{Majorana supersymmetries:}\quad &\xi^{*i\,\oa}=\xi_i^\ub\, B^{-1}{}_\ub{}^\oa,
  \label{i2-8} \\
  \mbox{symplectic Majorana supersymmetries:}\quad &\xi^{*i\,\oa}=-\xi_j^\ub\, B^{-1}{}_\ub{}^\oa\,\Omega^{*ji}
	\label{i2-9} 
 \end{align}
where $\xi^{*i\,\oa}$ denotes the conjugate-complex of $\xi^\ua_i$. For the matrix $B$ there are two different options in all cases exept for signatures $(0,2)$ and $(2,0)$ in $\Dim=2$ dimensions (see section 2.8 of \pap). The results will be formulated in such a way that they are valid for both choices of $B$ (see section 5.5 of \pap). Therefore it will not be necessary to fix a choice of $B$. Symplectic Majorana supersymmetries occur only for the signatures $(0,3)$ and $(3,0)$ in $\Dim=3$ dimensions and the matrix $\Om$ used in these cases will be specified in section \ref{3D.1}.

In all cases we shall use the following strategy to compute $\Hg(\fb)$: we first compute the cohomology groups explicitly in a particular spinor representation and then reformulate the result in an \so{t,\Dim-t}-covariant way (with \so{t,\Dim-t}-transformations as in section 2.6 of \pap) so that they become independent of the spinor representation. 

We shall use the notation $\sim$ for equivalence in $\Hg(\fb)$, i.e. for $\om_1,\om_2\in\Omg$ the notation $\om_1\sim \om_2$ means $\om_1-\om_2=\fb\om_3$ for some $\om_3\in\Omg$:
\begin{align}
  \om_1\sim \om_2\quad :\Leftrightarrow \quad \exists\,\om_3:\ \om_1-\om_2=\fb\om_3\quad (\om_1,\om_2,\om_3\in\Omg).
  \label{equiv}
\end{align}

Notation and conventions which are not explained here are as in \pap.

\section{\texorpdfstring{Primitive elements in $\Dim=2$ dimensions}{Primitive elements in D=2 dimensions}}\label{2D}

\subsection{\texorpdfstring{$\Hg(\fb)$ for signature (1,1) in a particular representation}{H(gh) for signature (1,1) in a particular representation}}\label{2D.1}

We shall first compute $\Hg(\fb)$ in $\Dim=2$ dimensions for signature $(1,1)$ and any numbers $\Ns_+,\Ns_-$ of Majorana-Weyl supersymmetries in a spinor representation with
 \begin{align}
  \gam^1=-\Ii\, \sigma_1\, ,\ \gam^2=\sigma_2\, ,\ \GAM=\sigma_3\, ,\ \CC=\sigma_2\, .
	\label{2D1} 
 \end{align}
In this spinor representation Majorana-Weyl supersymmetry ghosts $\xi^\pm_{i_\pm}=(\xi^{\pm\ul1}_{i_\pm},\xi^{\pm\ul2}_{i_\pm})$  (with $\xi^\pm_{i_\pm}\GAM=\pm\xi^\pm_{i_\pm}$) have only one nonvanishing component,
\begin{align}
  \xi^+_{i_+}=(\psi_{i_+},0),\quad 
  \xi^-_{i_-}=(0,\Ii\chi_{i_-}).
	\label{2D3} 
\end{align}
The coboundary operator $\fb$ acts on the translation ghosts according to
 \begin{align}
  \fb c^1=\ihalf\sum_{i_+=1}^{\Ns_+}(\psi_{i_+})^2
  +\ihalf\sum_{i_-=1}^{\Ns_-}(\chi_{i_-})^2,\quad \fb c^2=
  \ihalf\sum_{i_+=1}^{\Ns_+}(\psi_{i_+})^2
  -\ihalf\sum_{i_-=1}^{\Ns_-}(\chi_{i_-})^2.
	\label{2D4} 
 \end{align}
These transformations can be simplified by introducing the following purely imaginary linear combinations of the translation ghosts:
\begin{align}
  \tc1=-\Ii\,(c^1+c^2)\,,\quad \tc2=-\Ii\,(c^1-c^2).
	\label{2D5} 
\end{align}
$\tc1$ and $\tc2$ have the $\fb$-transformations
\begin{align}
  \fb\tc1=\sum_{i_+=1}^{\Ns_+}(\psi_{i_+})^2\,,\quad \fb\tc2=\sum_{i_-=1}^{\Ns_-}(\chi_{i_-})^2 .
	\label{2D6} 
\end{align}

We define the space $\Om_+$ of polynomials $\om_+(\tc1,\psi_{1_+},\dots,\psi_{\Ns_+})$ in $\tc1$ and the components of the supersymmetry ghosts of positive chirality, and the space $\Om_-$ of polynomials $\om_-(\tc2,\chi_{1_-},\dots,\chi_{\Ns_-})$ in $\tc2$ and the components of the supersymmetry ghosts of negative chirality. 
Equation \eqref{2D6} shows that $\fb$ does not lead out of these spaces respectively, i.e. $\om_+\in\Om_+$ implies $(\fb\om_+)\in\Om_+$ and $\om_-\in\Om_-$ implies $(\fb\om_-)\in\Om_-$, for all $\om_+$ and $\om_-$. Furthermore the space $\Omg$ of polynomials in all ghost variables can be written as the tensor product $\Om_+\otimes\Om_-$ of $\Om_+$ and $\Om_-$ (with $\om_+\otimes \om_-=\om_+\om_-$). This implies the K\"unneth formula $\Hg(\fb)=H_+(\fb)\otimes H_-(\fb)$ where $H_+(\fb)$ and $H_-(\fb)$ denote the cohomology of $\fb$ in $\Om_+$ and $\Om_-$ respectively. 

$H_+(\fb)$ and $H_-(\fb)$ can be directly obtained from lemmas 6.1 and 6.2 of \pap. Indeed, up to a factor $\half$, $\fb\tc1$ and $\fb\tc2$ are completely analogous to $\fb c^1$ in $\Dim=1$ dimension, cf. equation (6.5) of \pap. We conclude that $H_+(\fb)$ is for $\Ns_+>0$ represented by polynomials $a_0(\psi_{2_+},\dots,\psi_{\Ns_+})+\psi_{1_+} a_1(\psi_{2_+},\dots,\psi_{\Ns_+})$ and that $H_-(\fb)$ is for $\Ns_->0$ represented by polynomials $b_0(\chi_{2_-},\dots,\chi_{\Ns_-})+\chi_{1_-} b_1(\chi_{2_-},\dots,\chi_{\Ns_-})$, where $a_0(\psi_{2_+},\dots,\psi_{\Ns_+})$ and $a_1(\psi_{2_+},\dots,\psi_{\Ns_+})$ are arbitrary polynomials in $\psi_{2_+},\dots,\psi_{\Ns_+}$ and $b_0(\chi_{2_-},\dots,\chi_{\Ns_-})$ and $b_1(\chi_{2_-},\dots,\chi_{\Ns_-})$ are arbitrary polynomials in $\chi_{2_-},\dots,\chi_{\Ns_-}$. The K\"unneth formula yields thus:

\begin{lemma}[$\Hg(\fb)$ for $\Ns_+>0$ and $\Ns_->0$]\label{lem2D2}\quad \\
In the spinor representation \eqref{2D1}, $\Hg(\fb)$ is represented in the cases with both $\Ns_+>0$ and $\Ns_->0$ by polynomials in the supersymmetry ghosts which are at most linear both in $\psi_{1_+}$ and in $\chi_{1_-}$ and do not depend on the translation ghosts:
\begin{align}
&\fb\om=0\ \Leftrightarrow\ \om\sim 
a_{00}+\psi_{1_+} a_{10}+\chi_{1_-} a_{01}+\psi_{1_+}\chi_{1_-} a_{11};
\label{2D10}\\
&a_{00}+\psi_{1_+} a_{10}+\chi_{1_-} a_{01}+\psi_{1_+}\chi_{1_-} a_{11}\sim 0\
\Leftrightarrow\ a_{00}=a_{10}=a_{01}=a_{11}=0
\label{2D11}
\end{align}
where $a_{00}$, $a_{10}$, $a_{01}$ and $a_{11}$ are polynomials in 
$\psi_{2_+}$,\dots,$\psi_{\Ns_+}$ or $\chi_{2_-}$,\dots,$\chi_{\Ns_-}$ or complex numbers:
\begin{align}
\Ns_+>1,\Ns_->1:\quad &a_{ij}=a_{ij}(\psi_{2_+},\dots,\psi_{\Ns_+},\chi_{2_-},\dots,\chi_{\Ns_-}),\quad i,j\in\{0,1\};
\label{2D12}\\
\Ns_+>1,\Ns_-=1:\quad &a_{ij}=a_{ij}(\psi_{2_+},\dots,\psi_{\Ns_+}),\quad i,j\in\{0,1\};
\label{2D13}\\
\Ns_+=1,\Ns_->1:\quad &a_{ij}=a_{ij}(\chi_{2_-},\dots,\chi_{\Ns_-}),\quad i,j\in\{0,1\};
\label{2D14}\\
\Ns_+=1,\Ns_-=1:\quad &a_{ij}\in\mathbb{C},\quad i,j\in\{0,1\}.
\label{2D15}
\end{align}
\end{lemma}

The cases $\Ns_+=0$ or $\Ns_-=0$ are even simpler. E.g., in the case $\Ns_-=0$ one has $\fb\tc2=0$ and $\Om_-=\{a+b\tc2|a,b\in\mathbb{C}\}$. Hence, in this case $H_-(\fb)$ coincides with $\Om_-$ and the K\"unneth formula gives:

\begin{lemma}[$\Hg(\fb)$ for $\Ns_+>0$ and $\Ns_-=0$]\label{lem2D1}\quad \\
In the spinor representation \eqref{2D1}, $\Hg(\fb)$ is represented in the cases with $\Ns_+>0$ and $\Ns_-=0$ by polynomials in the supersymmetry ghosts which are at most linear in $\psi_{1_+}$ and do not depend on $\tc1$:
\begin{align}
&\fb\om=0\ \Leftrightarrow\ \om\sim 
a_0+\psi_{1_+} a_1;
\label{2D8}\\
&a_0+\psi_{1_+} a_1\sim 0\
\Leftrightarrow\ a_0=a_1=0
\label{2D9}
\end{align}
where $a_0$ and $a_1$ are polynomials in $\psi_{2_+}$, \dots, $\psi_{\Ns_+}$ or the translation ghost variable $\tc2$:
\begin{align}
\Ns_+>1:&\quad a_i=a_{i0}(\psi_{2_+},\dots,\psi_{\Ns_+})+\tc2 a_{i1}(\psi_{2_+},\dots,\psi_{\Ns_+}),\quad i\in\{0,1\};
\label{2D16}\\
\Ns_+=1:&\quad a_i=a_{i0}+\tc2 a_{i1},\ a_{i0},a_{i1}\in\mathbb{C},\quad i\in\{0,1\}.
\label{2D17}
\end{align}
\end{lemma}

An analogous result holds for $\Ns_+=0$ and $\Ns_->0$, with the $\chi_{i_-}$ in place of the $\psi_{i_+}$ and $\tc1$ in place of $\tc2$.

Notice that for two or more Majorana-Weyl supersymmetries it makes a considerable difference for the cohomology whether or not all the supersymmetries have the same chirality. In particular, in the case $(\Ns_+,\Ns_-)=(1,1)$ lemma \ref{lem2D2} states that $\Hg(\fb)$ is represented by $a_{00}+\psi_{1_+} a_{10}+\chi_{1_-} a_{01}+\psi_{1_+}\chi_{1_-} a_{11}$ with $a_{ij}\in\mathbb{C}$. Hence, $\Hg(\fb)$ is four dimensional in the case $(\Ns_+,\Ns_-)=(1,1)$ (counting complex dimensions). This differs from the case $(\Ns_+,\Ns_-)=(2,0)$ for which, according to lemma \ref{lem2D1}, $\Hg(\fb)$ is represented by $a_0(\psi_{2_+},\tc2)+\psi_{1_+} a_1(\psi_{2_+},\tc2)$ where $a_0(\psi_{2_+},\tc2)$ and $a_1(\psi_{2_+},\tc2)$ are polynomials of arbitrary degree in $\psi_{2_+}$ and may also depend linearly on $\tc2$. Hence, in the case $(\Ns_+,\Ns_-)=(2,0)$ the cohomology $\Hg(\fb)$ is infinite dimensional, in sharp contrast to the case $(\Ns_+,\Ns_-)=(1,1)$ which has the same number of supersymmetries.

\subsection{\texorpdfstring{$\Hg(\fb)$ for signature (1,1) in covariant form}{H(gh) for signature (1,1) in covariant form}}\label{2D.2}

The results summarized in lemmas \ref{lem2D2} and \ref{lem2D1} can be readily rewritten for spinor representations equivalent to the spinor representation \eqref{2D1}, using that the equivalence transformations relating any two spinor representations in even dimensions do not mix chiralities, cf. section 2.7 of \pap. Since the $\psi_{i_+}$ and $\chi_{i_-}$ denote the components of chiral supersymmetry ghosts in the spinor representation \eqref{2D1}, we can simply substitute the components of chiral supersymmetry ghosts in any equivalent spinor representation for them to obtain $\Hg(\fb)$ in the respective spinor representation. Furthermore, one readily checks that the product $\psi_{1_+}\chi_{1_-}$ can be written as the \so{1,1}-invariant $\xi^{+\ua}_{1_+} \xi^{-\ub}_{1_-}(\GAM\CC^{-1})_{\ua\ub}$ which extends it to spinor representations equivalent to \eqref{2D1}.
Therefore, a spinor representation independent formulation of lemma \ref{lem2D2} is, for instance:

\begin{lemma}[$\Hg(\fb)$ for $\Ns_+>0$ and $\Ns_->0$]\label{lem2D3}\quad \\
$\Hg(\fb)$ is represented in the cases with both $\Ns_+>0$ and $\Ns_->0$ by polynomials in the supersymmetry ghosts which are at most linear both in the components of $\xi_{1_+}^+$ and in the components of $\xi_{1_-}^-$ and do not depend on the translation ghosts:
\begin{align}
&\fb\om=0\ \Leftrightarrow\ \om\sim 
a+\xi_{1_+}^{+\ua}a_{+\ua}+\xi_{1_-}^{-\ua}a_{-\ua}+\xi^{+\ua}_{1_+} \xi^{-\ub}_{1_-}(\GAM\CC^{-1})_{\ua\ub}\,a_{+-}\,;
\label{2D18}\\
&a+\xi_{1_+}^{+\ua}a_{+\ua}+\xi_{1_-}^{-\ua}a_{-\ua}+\xi^{+\ua}_{1_+} \xi^{-\ub}_{1_-}(\GAM\CC^{-1})_{\ua\ub}\,a_{+-}\sim 0\notag\\
&\quad
\Leftrightarrow\quad a=\xi_{1_+}^{+\ua}a_{+\ua}=\xi_{1_-}^{-\ua}a_{-\ua}=a_{+-}=0
\label{2D19}
\end{align}
where $a$, $a_{+\ua}$, $a_{-\ua}$ and $a_{+-}$ are polynomials in the components of the supersymmetry ghosts
$\xi_{2_+}^+$,\dots,$\xi_{\Ns_+}^+$ or $\xi_{2_-}^-$,\dots,$\xi_{\Ns_-}^-$ or complex numbers analogously to equations \eqref{2D12} to \eqref{2D15}.
\end{lemma}

Analogously one may formulate lemma \ref{lem2D1} in a spinor representation independent form. Equation \eqref{2D19} takes into account that, in general, in a spinor representation different from (but equivalent to) the spinor representation \eqref{2D1} the nonvanishing components of $\xi_{1_+}^+$ and of $\xi_{1_-}^-$ are linearly dependent, respectively.

\subsection{\texorpdfstring{$\Hg(\fb)$ for signatures (0,2) and (2,0)}{H(gh) for signatures (0,2) and (2,0)}}\label{2D.3}

We now derive $\Hg(\fb)$ for signatures $(0,2)$ and $(2,0)$ departing from particular spinor representations with
 \begin{align}
  \mathrm{signature}\ (0,2):&\quad\gam^1=\sigma_1\, ,\ \gam^2=\sigma_2\, ,\ \GAM=\sigma_3\, ,\ 
  \CC=\sigma_2\, ;
	\label{2D20} \\
	\mathrm{signature}\ (2,0):&\quad\gam^1=-\Ii\,\sigma_1\, ,\ \gam^2=-\Ii\,\sigma_2\, ,\ \GAM=\sigma_3\, ,\ 
  \CC=\sigma_2\, .
	\label{2D20a}
 \end{align}
$\fb$ acts on the translation ghosts according to
 \begin{align}
  \mathrm{signature}\ (0,2):&\ \fb c^1=\half\sum_{i=1}^\Ns(-\xx1_i\xx1_i+\xx2_i\xx2_i),\
  \fb c^2=\ihalf\sum_{i=1}^\Ns(\xx1_i\xx1_i+\xx2_i\xx2_i);
	\label{2D21}\\
	\mathrm{signature}\ (2,0):&\ \fb c^1=\ihalf\sum_{i=1}^\Ns(\xx1_i\xx1_i-\xx2_i\xx2_i),\
  \fb c^2=\half\sum_{i=1}^\Ns(\xx1_i\xx1_i+\xx2_i\xx2_i).
  \label{2D21a}
 \end{align}
In terms of the ghost variables $\psi_i=\xx1_i$, $\chi_i=\xx2_i$ and
\begin{align}
  \mathrm{signature}\ (0,2):&\quad\tc1=-c^1-\Ii\, c^2,\quad \tc2=c^1-\Ii\, c^2;
	\label{2D22}\\
	\mathrm{signature}\ (2,0):&\quad\tc1=-\Ii\,c^1+c^2,\quad \tc2=\Ii\,c^1+c^2
	\label{2D22a}	 
\end{align}
the $\fb$-transformations \eqref{2D21} read in either case
\begin{align}
  \fb\tc1=\sum_{i=1}^{\Ns}(\psi_i)^2\,,\quad \fb\tc2=\sum_{i=1}^{\Ns}(\chi_i)^2 .
	\label{2D23} 
\end{align}
These transformations are analogous to those in equation
\eqref{2D6} for $\Ns_+=\Ns_-=\Ns$. Therefore we can directly obtain the cohomology $\Hg(\fb)$ for signature $(0,2)$ in the spinor representation \eqref{2D20} and for signature $(2,0)$ in the spinor representation \eqref{2D20a} from lemma \ref{lem2D2} for $\Ns_+=\Ns_-=\Ns$:

\begin{lemma}[$\Hg(\fb)$ in the particular spinor representations]\label{lem2D4}\quad \\
In the spinor representations \eqref{2D20} for signature $(0,2)$ and \eqref{2D20a} for signature $(2,0)$, $\Hg(\fb)$ is represented by polynomials in the supersymmetry ghosts which are at most linear both in $\psi_{1}$ and $\chi_{1}$ and do not depend on the translation ghosts:
\begin{align}
&\fb\om=0\ \Leftrightarrow\ \om\sim 
a_{00}+\psi_{1} a_{10}+\chi_{1} a_{01}+\psi_{1}\chi_{1} a_{11};
\label{2D24}\\
&a_{00}+\psi_{1} a_{10}+\chi_{1} a_{01}+\psi_{1}\chi_{1} a_{11}\sim 0\
\Leftrightarrow\ a_{00}=a_{10}=a_{01}=a_{11}=0
\label{2D25}
\end{align}
where $a_{00}$, $a_{10}$, $a_{01}$ and $a_{11}$ are polynomials in
$\psi_{2}$,\dots,$\psi_{\Ns}$,$\chi_{2}$,\dots,$\chi_{\Ns}$ or complex numbers:
\begin{align}
\Ns>1:\quad &a_{ij}=a_{ij}(\psi_{2},\dots,\psi_{\Ns},\chi_{2},\dots,\chi_{\Ns}),\quad i,j\in\{0,1\};
\label{2D26}\\
\Ns=1:\quad &a_{ij}\in\mathbb{C},\quad i,j\in\{0,1\}.
\label{2D27}
\end{align}
\end{lemma}

To formulate $\Hg(\fb)$ in spinor representations equivalent to the spinor representations \eqref{2D20} and \eqref{2D20a} we use that $\psi_1$ and $\chi_1$ are the components of the supersymmetry ghost $\xi_1$ and that the product $\psi_{1}\chi_{1}$ equals the \so{t,2-t} invariant $\ihalf\xi^{\ua}_{1} \xi^{\ub}_{1}(\GAM\CC^{-1})_{\ua\ub}$ in these representations. This yields:

\begin{lemma}[$\Hg(\fb)$ in covariant form]\label{lem2D5}\quad \\
$\Hg(\fb)$ is for signatures $(0,2)$ and $(2,0)$ represented by cocycles $a$, $\xi_{1}^{\ua}a_{\ua}$ and\\ $\xi^{\ua}_{1} \xi^{\ub}_{1}(\GAM\CC^{-1})_{\ua\ub}\,a_{+-}$ where $a$, $a_{\ua}$ and $a_{+-}$ are polynomials in the components of the supersymmetry ghosts
$\xi_{2}$,\dots,$\xi_{\Ns}$ (if $\Ns>1$) or complex numbers (if $\Ns=1$) analogously to equations \eqref{2D26} and \eqref{2D27}:
\begin{align}
&\fb\om=0\ \Leftrightarrow\ \om\sim 
a+\xi_{1}^{\ua}a_{\ua}+\xi^{\ua}_{1} \xi^{\ub}_{1}(\GAM\CC^{-1})_{\ua\ub}\,a_{+-}\,;
\label{2D28}\\
&a+\xi_{1}^{\ua}a_{\ua}+\xi^{\ua}_{1} \xi^{\ub}_{1}(\GAM\CC^{-1})_{\ua\ub}\,a_{+-}\sim 0\notag\\
&\quad
\Leftrightarrow\quad a=a_{\ua}=a_{+-}=0.
\label{2D29}
\end{align}
\end{lemma}

\section{\texorpdfstring{Primitive elements in $\Dim=3$ dimensions}{Primitive elements in D=3 dimensions}}\label{3D}

\subsection{\texorpdfstring{$\Hg(\fb)$ in a particular representation}{H(gh) in a particular representation}}\label{3D.1}

In $\Dim=3$ dimensions we first compute $\Hg(\fb)$ for any signature $(t,3-t)$ in a particular spinor representation given by
 \begin{align}
 \gam_a=k_a\,\sigma_a\, ,\quad a\in\{1,2,3\},\quad 
 k_a= \left\{\begin{array}{rl}\Ii & \mbox{for\ $a\leq t$}\\
  1& \mbox{for\ $a>t$}\end{array}\right.,\,\quad \CC=\sigma_2\,.
  \label{3D1}
 \end{align}
The supersymmetry ghosts $\xi_i$ are for signatures $(1,2)$ and $(2,1)$ Majorana spinors fulfilling equation \eqref{i2-8} and for signatures $(0,3)$ and $(3,0)$ symplectic Majorana spinors fulfilling equation \eqref{i2-9} with a matrix $\Om$ given by 
 \begin{align}
  \Om=\begin{pmatrix} E & 0 & \cdots & 0\\ 
                      0 & E & \cdots & 0\\
                      \vdots & \vdots & \ddots & \vdots\\
                      0 & 0 & \cdots & E                 
      \end{pmatrix},\quad
  E=\begin{pmatrix} 0 & 1 \\ -1 & 0 \end{pmatrix}.
	\label{3D68} 
 \end{align}
Accordingly, for signatures $(0,3)$ and $(3,0)$ we consider even numbers $\Ns$ of sets of supersymmetries and supersymmetry ghosts.

We introduce the following notation for the components of the supersymmetry ghosts:
 \begin{align}
  (\xx1_i,\xx2_i)=(\psi_i,\Ii \chi_i),
	\label{3D2} 
 \end{align}
and the following translation ghost variables (with $k_a$ as in \eqref{3D1}):
\begin{align}
  \tc1=-k_1\,c^1-\Ii\,k_2\, c^2,\quad \tc2=-k_1 \,c^1+\Ii\, k_2\, c^2,\quad \tc3=-\Ii\,k_3\, c^3.
	\label{3D5} 
\end{align}
In terms of these ghost variables the coboundary operator $\fb$ acts for all signatures $(t,3-t)$ according to
\begin{align}
  \fb\tc1=\sum_{i=1}^{\Ns}(\psi_{i})^2,\quad \fb\tc2=\sum_{i=1}^{\Ns}(\chi_{i})^2,\quad
  \fb \tc3=\sum_{i=1}^{\Ns}\psi_{i}\chi_{i}\, .
	\label{3D6} 
\end{align}

\subsubsection{Strategy}\label{3D.1.1}

In order to compute $\Hg(\fb)$ in the spinor representation \eqref{3D1} we shall use results in $D=2$ dimensions obtained in section \ref{2D}.
To use these results we define the space $\7\Om$ of polynomials in the ghost variables that do not depend on $\tc3$,
\begin{align}
  \7\Om:=\left\{\7\om\in\Omg\ \Big|\ \frac{\6\7\om}{\6\tc3}=0\right\}.
	\label{3D7} 
\end{align}
The coboundary operator $\fb$ acts in the space $\7\Om$ exactly as on ghost polynomials in $\Dim=2$ dimensions for signatures $(0,2)$ and $(2,0)$, cf. equations \eqref{2D23}. The cohomology of $\fb$ in $\7\Om$ is thus obtained from lemma \ref{lem2D4}. We denote this cohomology by $\hHg(\fb)$.

To determine $\Hg(\fb)$ from $\hHg(\fb)$ we write a ghost polynomial $\om\in\Omg$ as
\begin{align}
  \om=\7\om_0+\tc3\7\om_1\, ,\quad \7\om_0,\7\om_1\in\7\Om.
	\label{3D8} 
\end{align}
This yields
\begin{align}
  \fb\om=\fb\7\om_0+\sum_{i=1}^{\Ns}\psi_{i}\chi_{i}\7\om_1-\tc3(\fb\7\om_1).
	\label{3D9} 
\end{align}
Notice that on the right hand side of equation \eqref{3D9} only the last term contains $\tc3$. We thus obtain:
\begin{align}
  \fb\om=0\quad\LRA\quad 
  \fb\7\om_1=0\ \wedge\
   \fb\7\om_0+\sum_{i=1}^{\Ns}\psi_{i}\chi_{i}\7\om_1=0.
  \label{3D10}
\end{align}
The first condition $\fb\7\om_1=0$ in \eqref{3D10} imposes that $\7\om_1$ is a cocycle in $\hHg(\fb)$. This condition will be solved by means of the result \eqref{2D24} of lemma \ref{lem2D4}. The second condition in \eqref{3D10} imposes that $\sum_{i=1}^{\Ns}\psi_{i}\chi_{i}\7\om_1$ is a coboundary in $\hHg(\fb)$. That second condition will be solved by means of the result \eqref{2D25} of lemma \ref{lem2D4}. Then $\7\om_0$ will be determined using again the result \eqref{2D24} of lemma \ref{lem2D4}, and $\om$ will be obtained from the results for $\7\om_0$ and $\7\om_1$ using \eqref{3D8}. We shall have to distinguish the cases $\Ns=1$ and $\Ns>1$.

\subsubsection{\texorpdfstring{$\Hg(\fb)$ for $\Ns=1$}{H(gh) for N=1}}\label{3D.1.2}

Starting from the first condition $\fb\7\om_1=0$ in \eqref{3D10} we conclude in the case $\Ns=1$ from
the result \eqref{2D24} of lemma \ref{lem2D4} that
\begin{align}
  \7\om_1=\fb\7\varrho_1+a_{00}+\psi_1 a_{10}+\chi_1 a_{01}+\psi_1\chi_1 a_{11}
  \label{3D11}
\end{align}
for some polynomial $\7\varrho_1\in\7\Om$ and some complex numbers $a_{ij}\in\mathbb{C}$. Using this result for $\7\om_1$ in the second condition in \eqref{3D10}, the latter becomes in the case $\Ns=1$:
\begin{align}
  0
   &=\fb\7\om_0+\psi_1\chi_1(\fb\7\varrho_1+a_{00}+\psi_1 a_{10}+\chi_1 a_{01}+\psi_1\chi_1 a_{11})\notag\\
   &=\fb(\7\om_0+\psi_1\chi_1\7\varrho_1+\tc1\chi_1 a_{10}+\tc2\psi_1 a_{01}
         +\half (\tc1\chi_1\chi_1+\tc2\psi_1\psi_1) a_{11})+\psi_1\chi_1 a_{00}\, .
  \label{3D12}
\end{align}
Equation \eqref{3D12} imposes in particular that $\psi_1\chi_1 a_{00}$ is a coboundary in $\hHg(\fb)$. Using the result \eqref{2D25} of lemma \ref{lem2D4} we conclude
\begin{align}
  a_{00}=0.
  \label{3D13}
\end{align}
Using now equation \eqref{3D13} in equation \eqref{3D12}, the latter imposes
\[
\fb(\7\om_0+\psi_1\chi_1\7\varrho_1+\tc1\chi_1 a_{10}+\tc2\psi_1 a_{01}
         +\half (\tc1\chi_1\chi_1+\tc2\psi_1\psi_1)a_{11})=0.
\]
Using again the first result \eqref{2D24} of lemma \ref{lem2D4}, we conclude
\begin{align}
  &\7\om_0+\psi_1\chi_1\7\varrho_1+\tc1\chi_1 a_{10}+\tc2\psi_1 a_{01}
         +\half (\tc1\chi_1\chi_1+\tc2\psi_1\psi_1)a_{11}\notag\\
  &=\fb\7\varrho_0+b_{00}+\psi_1 b_{10}+\chi_1 b_{01}+\psi_1\chi_1 b_{11}
  \label{3D14}
\end{align}
for some $\7\varrho_0\in\7\Om$ and some $b_{ij}\in\mathbb{C}$. Solving equation \eqref{3D14} for $\7\om_0$ and using the results for $\7\om_0$ and $\7\om_1$ in equation \eqref{3D8} we obtain
\begin{align}
  \om 
  =&-\psi_1\chi_1\7\varrho_1-\tc1\chi_1 a_{10}-\tc2\psi_1 a_{01}
         -\half (\tc1\chi_1\chi_1+\tc2\psi_1\psi_1)a_{11}\notag\\
  &+\fb\7\varrho_0+b_{00}+\psi_1 b_{10}+\chi_1 b_{01}+\psi_1\chi_1 b_{11}\notag\\
  &+\tc3(\fb\7\varrho_1+\psi_1 a_{10}+\chi_1 a_{01}+\psi_1\chi_1 a_{11})\notag\\
  =&\,\fb(\7\varrho_0-\tc3\7\varrho_1+\tc3 b_{11})+b_{00}+\psi_1 b_{10}+\chi_1 b_{01}
  +(\tc3\psi_1-\tc1\chi_1)a_{10}\notag\\
  &+(\tc3\chi_1-\tc2\psi_1)a_{01}
  +(\tc3\psi_1\chi_1-\half \tc1\chi_1\chi_1-\half\tc2\psi_1\psi_1)a_{11}\, .
  \label{3D15}
\end{align}
The cocyles $b_{00}+\psi_1 b_{10}+\chi_1 b_{01}
  +(\tc3\psi_1-\tc1\chi_1)a_{10}
  +(\tc3\chi_1-\tc2\psi_1)a_{01}$ are at most linear in the supersymmetry ghosts and, therefore, cannot be exact in $\Hg(\fb)$ since $\fb$-coboundaries in $\Omg$ depend at least quadratically on the supersymmetry ghosts owing to equations \eqref{3D6}. Furthermore it can be readily checked explicitly that
$\tc3\psi_1\chi_1-\half \tc1\chi_1\chi_1-\half\tc2\psi_1\psi_1$ is not exact in $\Hg(\fb)$: in order to be a coboundary it would have to be of the form $\fb(d_{ab}c^a c^b)$ for some $d_{ab}\in\mathbb{C}$ but no such $d_{ab}$ exist. One can conclude the non-existence of the $d_{ab}$ without any calculation, using that $\tc3\psi_1\chi_1-\half \tc1\chi_1\chi_1-\half\tc2\psi_1\psi_1$ is actually an \so{t,3-t}-invariant ghost polynomial, cf. section \ref{3D.2}, and therefore, owing to the \so{t,3-t}-invariance of $\fb$, $d_{ab}c^a c^b$ would have to be \so{t,3-t}-invariant too; however, there is no nonvanishing \so{t,3-t}-invariant bilinear polynomial in the translation ghosts in dimensions $\Dim\geq 3$ (the only candidate bilinear polynomial would be proportional to $\eta_{ab}c^ac^b$ but this vanishes as the translation ghosts anticommute). We conclude:

\begin{lemma}[$\Hg(\fb)$ for $\Ns=1$]\label{lem3D1}\quad \\
In the spinor representation \eqref{3D1}, $\Hg(\fb)$ is in the case $\Ns=1$ represented by the cocycles $1$, $\psi_1$, $\chi_1$, $\tc3\psi_1-\tc1\chi_1$, $\tc3\chi_1-\tc2\psi_1$ and $\tc3\psi_1\chi_1-\half \tc1\chi_1\chi_1-\half\tc2\psi_1\psi_1$:
\begin{align}
  &\fb\om=0\ \LRA\ \om\sim b_{00}+\psi_1 b_{10}+\chi_1 b_{01}
  +(\tc3\psi_1-\tc1\chi_1)a_{10}
  +(\tc3\chi_1-\tc2\psi_1)a_{01}\notag\\
  &\phantom{\fb\om=0\ \LRA\ \om\sim }+(\tc3\psi_1\chi_1-\half \tc1\chi_1\chi_1-\half\tc2\psi_1\psi_1)a_{11}\, ;
  \label{3D16}\\
  &b_{00}+\psi_1 b_{10}+\chi_1 b_{01}
  +(\tc3\psi_1-\tc1\chi_1)a_{10}
  +(\tc3\chi_1-\tc2\psi_1)a_{01}\notag\\
  &+(\tc3\psi_1\chi_1-\half \tc1\chi_1\chi_1-\half\tc2\psi_1\psi_1)a_{11}\sim 0\ \LRA\ b_{ij}=a_{ij}=0,\label{3D16a}
\end{align}
where $b_{ij},a_{ij}\in\mathbb{C}$.
\end{lemma}

\subsubsection{\texorpdfstring{Towards $\Hg(\fb)$ for $\Ns>1$}{Towards H(gh) for N>1}}\label{3D.1.3}

In the cases $\Ns>1$ we start again from the first condition $\fb\7\om_1=0$ in \eqref{3D10}. We conclude from
the result \eqref{2D24} for $\Ns>1$ in lemma \ref{lem2D4} that
\begin{align}
  \7\om_1=\fb\7\varrho_1+a_{00}+\psi_1 a_{10}+\chi_1 a_{01}+\psi_1\chi_1 a_{11}
  \label{3D23}
\end{align}
for some polynomial $\7\varrho_1\in\7\Om$, with $a_{ij}$ polynomials in $\psi_2,\dots,\psi_{\Ns},\chi_2,\dots,\chi_{\Ns}$:
\begin{align}
a_{ij}=a_{ij}(\psi_2,\dots,\psi_{\Ns},\chi_2,\dots,\chi_{\Ns}),\quad i,j\in\{0,1\}.
  \label{3D24}
\end{align}
Using this result for $\7\om_1$ in the second condition in \eqref{3D10}, the latter yields in the cases $\Ns>1$:
\begin{align}
  0
   =&\,\fb\7\om_0+\Sigma_i\psi_i\chi_i(\fb\7\varrho_1+a_{00}+\psi_1 a_{10}+\chi_1 a_{01}+\psi_1\chi_1 a_{11})
     \notag\\
   =&\,\fb(\7\om_0+\Sigma_i\psi_i\chi_i\7\varrho_1)
     +\psi_1\chi_1(a_{00}+\psi_1 a_{10}+\chi_1 a_{01}+\psi_1\chi_1 a_{11})\notag\\
     &+\Sigma'_i\psi_i\chi_i(a_{00}+\psi_1 a_{10}+\chi_1 a_{01}+\psi_1\chi_1 a_{11})\notag\\     
   =&\,\fb(\7\om_0+\Sigma_i\psi_i\chi_i\7\varrho_1+\tc1\chi_1 a_{10}+\tc2\psi_1 a_{01}
     +\half (\tc1\chi_1\chi_1+\tc2\psi_1\psi_1) a_{11})\notag\\
    &+\psi_1\chi_1 a_{00}-\Sigma'_i(\psi_i\psi_i\chi_1 a_{10}+\chi_i\chi_i\psi_1 a_{01})
     -\half \Sigma'_i(\psi_i\psi_i\chi_1\chi_1+\chi_i\chi_i\psi_1\psi_1) a_{11}\notag\\   
    &+\Sigma'_i\psi_i\chi_i(a_{00}+\psi_1 a_{10}+\chi_1 a_{01}+\psi_1\chi_1 a_{11})\notag\\  
   =&\,\fb(\7\om_0+\Sigma_i\psi_i\chi_i\7\varrho_1+\tc1\chi_1 a_{10}+\tc2\psi_1 a_{01}\notag\\  
     &+\half (\tc1\chi_1\chi_1+\tc2\psi_1\psi_1
     -\Sigma'_i\tc2\psi_i\psi_i-\Sigma'_i\tc1\chi_i\chi_i)a_{11})\notag\\
    &+\psi_1\chi_1 a_{00}-\Sigma'_i(\psi_i\psi_i\chi_1 a_{10}+\chi_i\chi_i\psi_1 a_{01})
     +\Sigma'_i\psi_i\psi_i\Sigma'_j\chi_j\chi_j a_{11}\notag\\   
    &+\Sigma'_i\psi_i\chi_i(a_{00}+\psi_1 a_{10}+\chi_1 a_{01}+\psi_1\chi_1 a_{11})
  \label{3D25}
\end{align}
where we used the notation
\[
\Sigma_i:=\sum_{i=1}^\Ns\, ,\quad \Sigma'_i:=\sum_{i=2}^\Ns\, .
\]
Equation \eqref{3D25} imposes:
\begin{align}
  \fb\7\varrho_2=
    &\,\psi_1\chi_1 (a_{00}+\Sigma'_i\psi_i\chi_i a_{11})
    +\Sigma'_i\psi_i\chi_i a_{00}+\Sigma'_i\psi_i\psi_i\Sigma'_j\chi_j\chi_j a_{11}\notag\\
    &+\psi_1\Sigma'_i(\psi_i\chi_i a_{10}-\chi_i\chi_i a_{01})
    +\chi_1\Sigma'_i(\psi_i\chi_i a_{01}-\psi_i\psi_i a_{10})
  \label{3D26}
\end{align}
for $\7\varrho_2=(-\7\om_0+\dots)\in\7\Om$.
Using the result \eqref{2D25} we conclude from equation \eqref{3D26} that
\begin{align}
  &a_{00}+\Sigma'_i\psi_i\chi_i a_{11}=0,\quad
  \Sigma'_i\psi_i\chi_i a_{00}+\Sigma'_i\psi_i\psi_i\Sigma'_j\chi_j\chi_j a_{11}=0,\notag\\ 
  &\Sigma'_i(\psi_i\chi_i a_{10}-\chi_i\chi_i a_{01})=0,\quad
  \Sigma'_i(\psi_i\chi_i a_{01}-\psi_i\psi_i a_{10})=0.
  \label{3D27}
\end{align}
The first and the second of these conditions imply
\begin{align}
  (\Sigma'_i\psi_i\chi_i \Sigma'_j\psi_j\chi_j-\Sigma'_i\psi_i\psi_i\Sigma'_j\chi_j\chi_j) a_{11}=0 
  \label{3D28}
\end{align}
which holds identically (for any $a_{11}$) in the case $\Ns=2$ and imposes $a_{11}=0$ in the cases $\Ns>2$. 

The third and the fourth of the conditions in \eqref{3D27} give:
\begin{align}
  \Ns=2:\quad &\psi_2 a_{10}-\chi_2 a_{01}=0;\label{3D27a}\\
  \Ns>2:\quad &a_{10}(\Sigma'_i\psi_i\chi_i \Sigma'_j\psi_j\chi_j-\Sigma'_i\psi_i\psi_i\Sigma'_j\chi_j\chi_j)=0,\notag\\ 
  &a_{01}(\Sigma'_i\psi_i\chi_i \Sigma'_j\psi_j\chi_j-\Sigma'_i\psi_i\psi_i\Sigma'_j\chi_j\chi_j)=0.
  \label{3D27b}
\end{align}
\eqref{3D27a} and \eqref{3D27b} imply
\begin{align}
  \Ns=2:\quad &a_{10}=\chi_2 b,\quad a_{01}=\psi_2 b;\label{3D27c}\\
  \Ns>2:\quad &a_{10}=a_{01}=0
  \label{3D27d}
\end{align}
for some polynomial $b$ in $\psi_2$ and $\chi_2$.

We thus infer from
\eqref{3D27}:
\begin{align}
  \Ns=2:&\quad  a_{00}=-\psi_2\chi_2 a_{11}\, ,\quad a_{10}=\chi_2 b\, ,\quad a_{01}=\psi_2 b\, ;
  \label{3D29}\\
  \Ns>2:&\quad  a_{00}=a_{11}=a_{10}=a_{01}=0
  \label{3D30}
\end{align}
where in \eqref{3D29} $a_{11}=a_{11}(\psi_2,\chi_2)$ and $b=b(\psi_2,\chi_2)$ are polynomials in $\psi_2$ and $\chi_2$ which are not constrained by the cocycle condition.
To proceed, we have to distinguish the cases $\Ns=2$ and $\Ns>2$.

\subsubsection{\texorpdfstring{$\Hg(\fb)$ for $\Ns=2$}{H(gh) for N=2}}\label{3D.1.4}

Using \eqref{3D29} in equations \eqref{3D23} and \eqref{3D25} we obtain
\begin{align}
  &\7\om_1=\fb\7\varrho_1+(\psi_1\chi_2+\chi_1\psi_2)b+(\psi_1\chi_1-\psi_2\chi_2) a_{11}\, ;
  \label{3D32}\\
  &\fb(\7\om_0+\Sigma_i\psi_i\chi_i\7\varrho_1+(\tc1\chi_1\chi_2+\tc2\psi_1\psi_2)b
  \notag\\
  &\phantom{\fb(}+\half (\tc1\chi_1\chi_1+\tc2\psi_1\psi_1
     -\tc2\psi_2\psi_2-\tc1\chi_2\chi_2)a_{11})=0.
  \label{3D33}
\end{align}
Using once again the result \eqref{2D24} of lemma \ref{lem2D4} we infer from \eqref{3D33}:
\begin{align}
&\7\om_0+\Sigma_i\psi_i\chi_i\7\varrho_1+(\tc1\chi_1\chi_2+\tc2\psi_1\psi_2)b\notag\\
&+\half (\tc1(\chi_1\chi_1-\chi_2\chi_2)+\tc2(\psi_1\psi_1
     -\psi_2\psi_2))a_{11}\notag\\
   &=\fb\7\varrho_0+b_{00}+\psi_1 b_{10}+\chi_1 b_{01}+\psi_1\chi_1 b_{11}
  \label{3D34}
\end{align}
for some $\7\varrho_0\in\7\Om$ and some polynomials $b_{ij}=b_{ij}(\psi_2,\chi_2)$ in $\psi_2$ and $\chi_2$.
Using the results for $\7\om_0$ and $\7\om_1$ in equation \eqref{3D8}, we obtain:
\begin{align}
\om=&-\Sigma_i\psi_i\chi_i\7\varrho_1-(\tc1\chi_1\chi_2+\tc2\psi_1\psi_2)b\notag\\
    &-\half (\tc1(\chi_1\chi_1-\chi_2\chi_2)+\tc2(\psi_1\psi_1
     -\psi_2\psi_2))a_{11}\notag\\
   &+\fb\7\varrho_0+b_{00}+\psi_1 b_{10}+\chi_1 b_{01}+\psi_1\chi_1 b_{11}\notag\\
   &+\tc3(\fb\7\varrho_1+(\psi_1\chi_2+\chi_1\psi_2)b+(\psi_1\chi_1-\psi_2\chi_2) a_{11})\notag\\
   =&\,\fb(\7\varrho_0-\tc3\7\varrho_1+\tc3 b_{11})+b'_{00}+\psi_1 b_{10}+\chi_1 b_{01}\notag\\
   &+(\tc3(\psi_1\chi_2+\chi_1\psi_2)-\tc1\chi_1\chi_2-\tc2\psi_1\psi_2)b\notag\\
   &+(\tc3(\psi_1\chi_1-\psi_2\chi_2)-\half \tc1(\chi_1\chi_1-\chi_2\chi_2)-\half\tc2(\psi_1\psi_1
     -\psi_2\psi_2))a_{11}
  \label{3D35}
\end{align}
where $b'_{00}=b_{00}-\psi_2\chi_2 b_{11}$. 

We have thus shown:
\begin{align}
\fb\om=0\ \LRA\ \om\sim &\,b'_{00}+\psi_1 b_{10}+\chi_1 b_{01}
+(\tc3(\psi_1\chi_2+\chi_1\psi_2)-\tc1\chi_1\chi_2-\tc2\psi_1\psi_2)b\notag\\
&\,+(\tc3(\psi_1\chi_1-\psi_2\chi_2)-\half \tc1(\chi_1\chi_1-\chi_2\chi_2)-\half\tc2(\psi_1\psi_1
     -\psi_2\psi_2))a_{11}\, .
  \label{3D37}
\end{align}

We shall now investigate whether and which cocycles in equation \eqref{3D37} are coboundaries, i.e., we shall study the equation
\begin{align}
&\,b'_{00}+\psi_1 b_{10}+\chi_1 b_{01}
+(\tc3(\psi_1\chi_2+\chi_1\psi_2)-\tc1\chi_1\chi_2-\tc2\psi_1\psi_2)b\notag\\
&\,+(\tc3(\psi_1\chi_1-\psi_2\chi_2)-\half \tc1(\chi_1\chi_1-\chi_2\chi_2)-\half\tc2(\psi_1\psi_1
     -\psi_2\psi_2))a_{11}=\fb\varrho.
  \label{3D38}
\end{align}
As in equation \eqref{3D8} we write $\varrho$ as
\begin{align}
\varrho=\7\varrho_3+\tc3\7\varrho_4\, ,\quad \7\varrho_3,\7\varrho_4\in\7\Om.
  \label{3D39}
\end{align}
Writing $\fb\varrho$ analogously to $\fb\om$ in \eqref{3D9}, equation \eqref{3D38} becomes
\begin{align}
&\,b'_{00}+\psi_1 b_{10}+\chi_1 b_{01}
+(\tc3(\psi_1\chi_2+\chi_1\psi_2)-\tc1\chi_1\chi_2-\tc2\psi_1\psi_2)b\notag\\
&\,+(\tc3(\psi_1\chi_1-\psi_2\chi_2)-\half \tc1(\chi_1\chi_1-\chi_2\chi_2)-\half\tc2(\psi_1\psi_1
     -\psi_2\psi_2))a_{11}\notag\\
&\,=\fb\7\varrho_3+(\psi_1\chi_1+\psi_2\chi_2)\7\varrho_4-\tc3(\fb\7\varrho_4)\, .
  \label{3D40}
\end{align}
The terms in \eqref{3D40} containing $\tc3$ impose
\begin{align}
(\psi_1\chi_2+\chi_1\psi_2)b+(\psi_1\chi_1-\psi_2\chi_2)a_{11}=-\fb\7\varrho_4\, .
  \label{3D41}
\end{align}
Using the result \eqref{2D25} of lemma \ref{lem2D4}, we conclude from equation \eqref{3D41}
\begin{align}
&b=a_{11}=0;\label{3D42}\\
&\fb\7\varrho_4=0.\label{3D43}
\end{align}
Using the result \eqref{2D24} of lemma \ref{lem2D4}, we infer from \eqref{3D43}:
\begin{align}
\7\varrho_4=\fb\7\varrho'+d_{00}+\psi_1 d_{10}+\chi_1 d_{01}+\psi_1\chi_1 d_{11}\, ,\ \7\varrho'\in\7\Om\label{3D44}
\end{align}
with $d_{ij}=d_{ij}(\psi_2,\chi_2)$ polynomials in $\psi_2$ and $\chi_2$.

Using now \eqref{3D42} and \eqref{3D44} in equation \eqref{3D40}, the latter yields
\begin{align}
b'_{00}+\psi_1 b_{10}+\chi_1 b_{01}
=&\,\fb(\7\varrho_3+(\psi_1\chi_1+\psi_2\chi_2)\7\varrho')\notag\\
&+(\psi_1\chi_1+\psi_2\chi_2)(d_{00}+\psi_1 d_{10}+\chi_1 d_{01}+\psi_1\chi_1 d_{11})\notag\\
=&\,\fb(\7\varrho_3+(\psi_1\chi_1+\psi_2\chi_2)\7\varrho')\notag\\
&+(\psi_1\chi_1+\psi_2\chi_2)d_{00}\notag\\
&+\fb(\tc1\chi_1 d_{10})-\psi_2\psi_2\chi_1 d_{10}+\psi_2\chi_2\psi_1 d_{10}\notag\\
&+\fb(\tc2\psi_1 d_{01})-\chi_2\chi_2\psi_1 d_{01}+\psi_2\chi_2\chi_1 d_{01}\notag\\
&+\fb(\tc1\chi_1\chi_1 d_{11}-\tc2\psi_2\psi_2 d_{11})\notag\\
&+\psi_2\psi_2\chi_2\chi_2 d_{11}
+\psi_2\chi_2\psi_1\chi_1 d_{11}\, .
  \label{3D45}
\end{align}
This gives:
\begin{align}
&\, b'_{00}-\psi_2\chi_2 d_{00}-\psi_2\psi_2\chi_2\chi_2 d_{11}
+\psi_1 (b_{10}+\chi_2\chi_2 d_{01}-\psi_2\chi_2 d_{10})\notag\\
&+\chi_1(b_{01}+\psi_2\psi_2 d_{10}-\psi_2\chi_2 d_{01})
-\psi_1\chi_1 (d_{00}+\psi_2\chi_2 d_{11})=\fb\7\varrho_5
  \label{3D46}
\end{align}
for $\7\varrho_5=(\7\varrho_3+\dots)\in\7\Om$.
Using the result \eqref{2D25} of lemma \ref{lem2D4} we infer from \eqref{3D46}:
\begin{align}
&b'_{00}-\psi_2\chi_2 d_{00}-\psi_2\psi_2\chi_2\chi_2 d_{11}=0,\quad 
d_{00}+\psi_2\chi_2 d_{11}=0,
\label{3D47}\\
&b_{10}+\chi_2\chi_2 d_{01}-\psi_2\chi_2 d_{10}=0,\quad 
b_{01}+\psi_2\psi_2 d_{10}-\psi_2\chi_2 d_{01}=0.\label{3D48}
\end{align}
\eqref{3D47} implies $b'_{00}=0$, \eqref{3D48} provides those polynomials $b_{10}$, $b_{01}$ for which $\psi_1 b_{10}+\chi_1 b_{01}$ is a coboundary in $\Hg(\fb)$. The latter condition on $b_{10}$ and $b_{01}$ can be rewritten in terms of the cocycle $\psi_1 b_{10}+\chi_1 b_{01}$ as follows:
\begin{align}
 \psi_1 b_{10}+\chi_1 b_{01}&=(\psi_1 \psi_2\chi_2-\chi_1\psi_2\psi_2) d_{10}
 +(\chi_1\psi_2\chi_2-\psi_1\chi_2\chi_2)d_{01}\notag\\
 &=(\psi_1\chi_2-\chi_1\psi_2)(\psi_2d_{10}-\chi_2d_{01}).\label{3D48a}
\end{align}
We have thus shown:

\begin{lemma}[$\Hg(\fb)$ for $\Ns=2$]\label{lem3D2}\quad \\
In the spinor representation \eqref{3D1}, $\Hg(\fb)$ is in the case $\Ns=2$ represented by cocycles $b'_{00}$, $\psi_1b_{10}$, $\chi_1b_{01}$, $(\tc3(\psi_1\chi_2+\chi_1\psi_2)-\tc1\chi_1\chi_2-\tc2\psi_1\psi_2)b$ and
$(\tc3(\psi_1\chi_1-\psi_2\chi_2)-\half \tc1(\chi_1\chi_1-\chi_2\chi_2)-\half\tc2(\psi_1\psi_1
     -\psi_2\psi_2))a_{11}$ where $b'_{00}$, $b_{10}$, $b_{01}$, $b$ and $a_{11}$ are polynomials in $\psi_2$ and $\chi_2$:
\begin{align}
&\fb\om=0\ \Leftrightarrow\ \om\sim 
b'_{00}+\psi_1 b_{10}+\chi_1 b_{01}
+(\tc3(\psi_1\chi_2+\chi_1\psi_2)-\tc1\chi_1\chi_2-\tc2\psi_1\psi_2)b\notag\\
&\phantom{\fb\om=0\ \Leftrightarrow\ \om\sim }+(\tc3(\psi_1\chi_1-\psi_2\chi_2)-\half \tc1(\chi_1\chi_1-\chi_2\chi_2)-\half\tc2(\psi_1\psi_1
     -\psi_2\psi_2))a_{11}\,;
\label{3D51}\\
&b'_{00}+\psi_1 b_{10}+\chi_1 b_{01}
+(\tc3(\psi_1\chi_2+\chi_1\psi_2)-\tc1\chi_1\chi_2-\tc2\psi_1\psi_2)b\notag\\
&+(\tc3(\psi_1\chi_1-\psi_2\chi_2)-\half \tc1(\chi_1\chi_1-\chi_2\chi_2)-\half\tc2(\psi_1\psi_1
     -\psi_2\psi_2))a_{11}\sim 0 \notag\\
&\LRA\ b'_{00}=b=a_{11}=0\ \wedge\ 
\psi_1 b_{10}+\chi_1 b_{01}=
(\psi_1\chi_2-\chi_1\psi_2)(\psi_2d_{10}-\chi_2d_{01})\label{3D52}
\end{align}
for some polynomials $d_{10}$ and $d_{01}$ in $\psi_2$ and $\chi_2$.
\end{lemma}

\subsubsection{\texorpdfstring{$\Hg(\fb)$ for $\Ns>2$}{H(gh) for N>2}}\label{3D.1.5}

Using \eqref{3D30} in equations \eqref{3D23} and \eqref{3D25} we obtain
\begin{align}
  &\7\om_1=\fb\7\varrho_1\, ;
  \label{3D54}\\
  &\fb(\7\om_0+\Sigma_i\psi_i\chi_i\7\varrho_1)=0.
  \label{3D55}
\end{align}
Using once again the result \eqref{2D24} of lemma \ref{lem2D4} we conclude from \eqref{3D55}:
\begin{align}
\7\om_0+\Sigma_i\psi_i\chi_i\7\varrho_1=\fb\7\varrho_0+b_{00}+\psi_1 b_{10}+\chi_1 b_{01}+\psi_1\chi_1 b_{11}
  \label{3D56}
\end{align}
for some $\7\varrho_0\in\7\Om$ and some polynomials $b_{ij}=b_{ij}(\psi_2,\dots,\psi_\Ns,\chi_2,\dots,\chi_\Ns)$.
Using the results for $\7\om_0$ and $\7\om_1$ in equation \eqref{3D8}, we obtain:
\begin{align}
\om &=-\Sigma_i\psi_i\chi_i\7\varrho_1+\fb\7\varrho_0+b_{00}+\psi_1 b_{10}+\chi_1 b_{01}+\psi_1\chi_1 b_{11}
   +\tc3(\fb\7\varrho_1)\notag\\
   &=\fb(\7\varrho_0-\tc3\7\varrho_1+\tc3 b_{11})+b'_{00}+\psi_1 b_{10}+\chi_1 b_{01}
  \label{3D57}
\end{align}
with $b'_{00}=b_{00}-\Sigma'_i\psi_i\psi_i b_{11}$ with $\Sigma'_i=\sum_{i=2}^\Ns$.
We have thus shown in the cases $\Ns>2$:
\begin{align}
\fb\om=0\ \LRA\ \om\sim b'_{00}+\psi_1 b_{10}+\chi_1 b_{01}.
  \label{3D58}
\end{align}
We still have to determine those cocycles $b'_{00}+\psi_1 b_{10}+\chi_1 b_{01}$ that are coboundaries. Analogously to the case $\Ns=2$ we thus study the equation
\begin{align}
b'_{00}+\psi_1 b_{10}+\chi_1 b_{01}=\fb\varrho.
  \label{3D58a}
\end{align}
In the same way as in the analysis of equation \eqref{3D38} one derives the analog of \eqref{3D45} for $\Ns>2$:
\begin{align}
b'_{00}+\psi_1 b_{10}+\chi_1 b_{01}
=&\,\fb(\7\varrho_3+\Sigma_i\psi_i\chi_i\7\varrho')\notag\\
&+\Sigma_i\psi_i\chi_i(d_{00}+\psi_1 d_{10}+\chi_1 d_{01}+\psi_1\chi_1 d_{11})\notag\\
=&\,\fb(\7\varrho_3+\Sigma_i\psi_i\chi_i\7\varrho')\notag\\
&+\Sigma_i\psi_i\chi_id_{00}\notag\\
&+\fb(\tc1\chi_1 d_{10})-\Sigma'_i\psi_i\psi_i\chi_1 d_{10}+\Sigma'_i\psi_i\chi_i\psi_1 d_{10}\notag\\
&+\fb(\tc2\psi_1 d_{01})-\Sigma'_i\chi_i\chi_i\psi_1 d_{01}+\Sigma'_i\psi_i\chi_i\chi_1 d_{01}\notag\\
&+\fb(\tc1\chi_1\chi_1 d_{11}-\tc2\Sigma'_i\psi_i\psi_i d_{11})\notag\\
&+\Sigma'_i\psi_i\psi_i\Sigma'_j\chi_j\chi_j d_{11}
+\Sigma'_i\psi_i\chi_i\psi_1\chi_1 d_{11}\, .
  \label{3D59}
\end{align}
This yields:
\begin{align}
\fb\7\varrho_5=&\, b'_{00}-\Sigma'_i\psi_i\chi_i d_{00}-\Sigma'_i\psi_i\psi_i\Sigma'_j\chi_j\chi_j d_{11}\notag\\
&+\psi_1 (b_{10}+\Sigma'_i\chi_i\chi_i d_{01}-\Sigma'_i\psi_i\chi_i d_{10})\notag\\
&+\chi_1(b_{01}+\Sigma'_i\psi_i\psi_i d_{10}-\Sigma'_i\psi_i\chi_i d_{01})\notag\\
&-\psi_1\chi_1 (d_{00}+\Sigma'_i\psi_i\chi_i d_{11})
  \label{3D60}
\end{align}
with $\7\varrho_5=(\7\varrho_3+\dots)\in\7\Om$.
Using the result \eqref{2D25} of lemma \ref{lem2D4} we infer from \eqref{3D60}:
\begin{align}
&b'_{00}=\Sigma'_i\psi_i\chi_i d_{00}+\Sigma'_i\psi_i\psi_i\Sigma'_j\chi_j\chi_j d_{11},\quad 
d_{00}+\Sigma'_i\psi_i\chi_i d_{11}=0,
\label{3D61}\\
&b_{10}=-\Sigma'_i\chi_i\chi_i d_{01}+\Sigma'_i\psi_i\chi_i d_{10},\quad 
b_{01}=-\Sigma'_i\psi_i\psi_i d_{10}+\Sigma'_i\psi_i\chi_i d_{01}.
\label{3D62}
\end{align}
\eqref{3D61} implies 
\begin{align}
b'_{00}&=(-\Sigma'_i\psi_i\chi_i \Sigma'_j\psi_j\chi_j+\Sigma'_i\psi_i\psi_i\Sigma'_j\chi_j\chi_j) d_{11}\notag\\
&=\half\Sigma'_i\Sigma'_j(\psi_i\chi_j-\chi_i\psi_j)^2d_{11}\ .
\label{3D63}
\end{align}
\eqref{3D62} implies 
\begin{align}
\psi_1 b_{10}+\chi_1 b_{01}&=
\psi_1(-\Sigma'_i\chi_i\chi_i d_{01}+\Sigma'_i\psi_i\chi_i d_{10})
+\chi_1(-\Sigma'_i\psi_i\psi_i d_{10}+\Sigma'_i\psi_i\chi_i d_{01})\notag\\
&=\Sigma'_i\psi_i(\psi_1\chi_i-\chi_1\psi_i)d_{10}
-\Sigma'_i\chi_i(\psi_1\chi_i-\chi_1\psi_i)d_{01}\ .
\label{3D64}
\end{align}
This yields:

\begin{lemma}[$\Hg(\fb)$ for $\Ns>2$]\label{lem3D3}\quad \\
In the spinor representation \eqref{3D1}, $\Hg(\fb)$ is in the case $\Ns>2$ represented by cocycles $b'_{00}$, $\psi_1 b_{10}$ and $\chi_1 b_{01}$ where $b'_{00}$, $b_{10}$ and $b_{01}$ are polynomials in $\psi_2,\dots,\psi_\Ns,\chi_2,\dots,\chi_\Ns$:
\begin{align}
&\fb\om=0\ \Leftrightarrow\ \om\sim 
b'_{00}+\psi_1 b_{10}+\chi_1 b_{01}\,;\label{3D65}\\
&b'_{00}+\psi_1 b_{10}+\chi_1 b_{01}\sim 0\ 
\Leftrightarrow\ \notag\\
&b'_{00}=\half\sum_{i=2}^\Ns\sum_{j=2}^\Ns(\psi_i\chi_j-\chi_i\psi_j)^2d_{11}\ \wedge\ 
\notag\\
&\psi_1 b_{10}+\chi_1 b_{01}
=\sum_{i=2}^\Ns[\psi_i(\psi_1\chi_i-\chi_1\psi_i)d_{10}
-\chi_i(\psi_1\chi_i-\chi_1\psi_i)d_{01}]
\label{3D66}
\end{align}
for some polynomials $d_{11}$, $d_{10}$ and $d_{01}$ in $\psi_2,\dots,\psi_\Ns,\chi_2,\dots,\chi_\Ns$.
\end{lemma}

\subsection{\texorpdfstring{$\Hg(\fb)$ in covariant form}{H(gh) in covariant form}}\label{3D.2}

To derive a spinor representation independent formulation of lemmas \ref{lem3D1}, \ref{lem3D2} and \ref{lem3D3} we introduce the following \so{t,3-t}-covariant ghost polynomials:
\begin{align}
  \vartheta_i^\ua=c^a\xi^\ub_i\gam_{a\ub}{}^\ua,\quad \Theta_{ij}=c^a\xi^\ua_i\xi^\ub_j(\gam_a\IC)_{\ua\ub}\, .
  \label{3D20}
\end{align}
The $\Theta_{ij}$ fulfill:
\begin{align}
  \Theta_{ij}=\Theta_{ji}=\vartheta_i\cdot \xi_j=\vartheta_j\cdot \xi_i
  \label{3D20a}
\end{align}
where $\vartheta_i\cdot\xi_j$ denotes the \so{t,3-t}-invariant product of $\vartheta_i$ and $\xi_j$, cf. equation (2.39) of \pap.

In the spinor representation \eqref{3D1} one has, for all signatures $(t,3-t)$,
\begin{align}
  &\vartheta_i^{\underline{1}}=\Ii\,(\tc3\psi_i-\tc1\chi_i)\,,\quad
  \vartheta_i^{\underline{2}}=\tc3\chi_i-\tc2\psi_i\,,
  \label{3D17}\\
  &\Theta_{ij}=\Ii\,(\tc3(\psi_i\chi_j+\psi_j\chi_i)-\tc1\chi_i\chi_j-\tc2\psi_i\psi_j).
  \label{3D19}
\end{align}

The coboundary operator $\fb$ acts on $\vartheta_i$ and $\Theta_{ij}$ according to
\begin{align}
\fb\vartheta_i^\ua=\Ii\sum_{j=1}^\Ns(\xi_i\cdot\xi_j)\, \xi^\ua_j\, ,\quad
\fb\Theta_{ij}=-\Ii\sum_{k=1}^\Ns(\xi_i\cdot\xi_k)(\xi_j\cdot\xi_k).
\label{3D22a}
\end{align}
Equations \eqref{3D22a} can be easily verified explicitly in the spinor representation \eqref{3D1} using equations \eqref{3D17} and \eqref{3D19}. The validity of equations \eqref{3D22a} in the particular spinor representation implies their validity in any spinor representation equivalent to the particular spinor representation owing to their \so{t,3-t}-covariance.\footnote{Alternatively one may verify equations \eqref{3D22a} directly in a spinor representation independent manner using the "completeness relation" $\delta_\ua^\ub\delta_\ug^\ud+\gam^a{}_\ua{}^\ub\gam_{a\,\ug}{}^\ud=2\delta_\ua^\ud\delta_\ug^\ub$
of the gamma-matrices in $\Dim=3$ dimensions.}

Equations \eqref{3D17} and \eqref{3D19} show that the various ghost polynomials involving the translations ghosts which appear in lemmas \ref{lem3D1} and \ref{lem3D2} are proportional to ghost polynomials \eqref{3D20} expressed in the spinor representation \eqref{3D1} respectively. Using additionally $\psi_i=\xx1_i$, $\Ii\chi_i=\xx2_i$ and that $\psi_i\chi_j-\chi_i\psi_j$ is the \so{t,3-t}-invariant product $\xi_i\cdot\xi_j$ of $\xi_i$ and $\xi_j$ we can rewrite lemmas \ref{lem3D1}, \ref{lem3D2} and \ref{lem3D3} in an \so{t,3-t}-covariant form which extends them to all spinor representations equivalent to the spinor representation \eqref{3D1}.

Lemma \ref{lem3D1} yields:

\begin{lemma}[$\Hg(\fb)$ for $\Ns=1$]\label{lem3D1a}\quad \\
In the case $\Ns=1$ a complete set of independent cohomology classes of $\Hg(\fb)$ is $\{\,[1]$, $[\xi_1^{\underline{1}}]$, $[\xi_1^{\underline{2}}]$, $[\vartheta_1^{\underline{1}}]$, $[\vartheta_1^{\underline{2}}]$, $[\Theta_{11}]\,\}$ with $\vartheta_1^\ua$ and $\Theta_{11}$ as in equations \eqref{3D20}:
\begin{align}
&\fb\om=0\ \Leftrightarrow\ \om\sim 
a+ \xi_1^\ua a_\ua+\vartheta_1^\ua b_\ua+\Theta_{11}b\,;
\label{3D21}\\
&a+ \xi_1^\ua a_\ua+\vartheta_1^\ua b_\ua+\Theta_{11} b\sim 0\
\Leftrightarrow\ a=a_\ua=b_\ua=b=0
\label{3D22}
\end{align}
where $a,a_\ua,b_\ua,b\in\mathbb{C}$.
\end{lemma}

{\bf Comment:}
In the case $\Ns=1$ equations \eqref{3D22a} give $\fb\vartheta_1^\ua=0$ and $\fb\Theta_{11}=0$ owing to $\xi_1\cdot\xi_1=0$ (one has $\xi_i\cdot\xi_j=-\xi_j\cdot\xi_i$ in the present case). Note that lemma \ref{lem3D1a}
only applies to signatures $(1,2)$ and $(2,1)$ because $\Ns$ is even for signatures $(3,0)$ and $(0,3)$.

Lemma \ref{lem3D2} yields:

\begin{lemma}[$\Hg(\fb)$ for $\Ns=2$]\label{lem3D2a}\quad \\
In the case $\Ns=2$\\
(i) any cocycle in $\Hg(\fb)$ can be written as a polynomial in the components of the supersymmetry ghosts $\xi_1$, $\xi_2$ which is at most linear in the components of $\xi_1$, or as polynomials in the components of $\xi_2$ times $\Theta_{12}$ or $\Theta_{11}-\Theta_{22}$ with $\Theta_{ij}$ as in equations \eqref{3D20}:
\begin{align}
\fb\om=0\ \Leftrightarrow\ \om\sim 
a(\xi_2)+\xi_1^\ua a_\ua(\xi_2) +\Theta_{12}b_{1}(\xi_2)+(\Theta_{11}-\Theta_{22})b_2(\xi_2)
\label{3D51a}
\end{align}
where $a(\xi_2)$, $a_\ua(\xi_2)$, $b_1(\xi_2)$ and $b_2(\xi_2)$ are polynomials in the components of $\xi_2$;

(ii) a cocycle $a(\xi_2)+\xi_1^\ua a_\ua(\xi_2)+\Theta_{12}b_{1}(\xi_2) +(\Theta_{11}-\Theta_{22})b_2(\xi_2)$ is a coboundary in $\Hg(\fb)$ if and only if $a$, $b_1$ and $b_2$ vanish and if $\xi_1^\ua a_\ua(\xi_2) $ depends on $\xi_1$ only via the \so{t,3-t}-invariant product $\xi_1\cdot\xi_2$ and at least quadratically on the components of $\xi_2$:
\begin{align} 
&a(\xi_2)+\xi_1^\ua a_\ua(\xi_2) +\Theta_{12}b_{1}(\xi_2)+(\Theta_{11}-\Theta_{22})b_2(\xi_2)\sim 0\ 
\Leftrightarrow\ \notag\\
& a=b_1=b_2=0\ \wedge\ \xi_1^\ua a_\ua(\xi_2) =(\xi_1\cdot\xi_2)\, \xi^\ua_2 d_\ua(\xi_2)
\label{3D52a}
\end{align}
for some polynomials $d_\ua(\xi_2)$ in the components of $\xi_2$.
\end{lemma}

{\bf Comment:} In the case $\Ns=2$ equations \eqref{3D22a} yield
$\fb\vartheta_1^\ua=\Ii(\xi_1\cdot\xi_2) \xi^\ua_2$, $\fb\Theta_{12}=0$ and $\fb\Theta_{11}=\fb\Theta_{22}=-\Ii(\xi_1\cdot\xi_2)^2$.
These relations are behind the results that in \eqref{3D51a} there is no analog of the term  $\vartheta_1^\ua b_\ua$ in lemma \ref{lem3D1a}, that $(\xi_1\cdot\xi_2)\, \xi^\ua_2 d_\ua(\xi_2)$ is a coboundary in $\Hg(\fb)$ and that $\Theta_{12}$ and $\Theta_{11}-\Theta_{22}$ are cocycles in $\Hg(\fb)$ for $\Ns=2$.

Lemma \ref{lem3D3} yields:

\begin{lemma}[$\Hg(\fb)$ for $\Ns>2$]\label{lem3D3a}\quad \\
In the cases $\Ns>2$\\
(i) any cocycle in $\Hg(\fb)$ can be written as a polynomial in the components of the supersymmetry ghosts $\xi_1$,\dots,$\xi_\Ns$ which is at most linear in the components of $\xi_1$:
\begin{align}
\fb\om=0\ \Leftrightarrow\ \om\sim 
a(\xi_2,\dots,\xi_\Ns)+\xi_1^\ua a_\ua(\xi_2,\dots,\xi_\Ns)
\label{3D65a}
\end{align}
where $a(\xi_2,\dots,\xi_\Ns)$ and $a_\ua(\xi_2,\dots,\xi_\Ns)$ are polynomials in the components of $\xi_2$,\dots$,\xi_\Ns$;

(ii) a cocycle $a(\xi_2,\dots,\xi_\Ns)+\xi_1^\ua a_\ua(\xi_2,\dots,\xi_\Ns)$ is a coboundary in $\Hg(\fb)$ if and only if $a(\xi_2,\dots,\xi_\Ns)$ is proportional to the sum of all squared \so{t,3-t}-invariants $\xi_i\cdot\xi_j$ with $i,j\in\{2,\dots,\Ns\}$ and if
$\xi_1^\ua a_\ua(\xi_2,\dots,\xi_\Ns)$ depends on $\xi_1$ only via $\sum_{i=2}^\Ns(\xi_1\cdot\xi_i)\xx1_i$ or $\sum_{i=2}^\Ns(\xi_1\cdot\xi_i)\xx2_i$:
\begin{align} 
&a(\xi_2,\dots,\xi_\Ns)+\xi_1^\ua a_\ua(\xi_2,\dots,\xi_\Ns)\sim 0\ 
\Leftrightarrow\ \notag\\
& a(\xi_2,\dots,\xi_\Ns)=\half\sum_{i=2}^\Ns\sum_{j=2}^\Ns(\xi_i\cdot\xi_j)^2d(\xi_2,\dots,\xi_\Ns)\ \wedge\ 
\notag\\
&\xi_1^\ua a_\ua(\xi_2,\dots,\xi_\Ns)=\sum_{i=2}^\Ns(\xi_1\cdot\xi_i)\xi^\ua_id_\ua(\xi_2,\dots,\xi_\Ns)
\label{3D66a}
\end{align}
for some polynomials $d(\xi_2,\dots,\xi_\Ns)$ and $d_\ua(\xi_2,\dots,\xi_\Ns)$ in the components of $\xi_2$,\dots,$\xi_\Ns$.
\end{lemma}

{\bf Comment:} Equations \eqref{3D22a} imply
\[
\fb(\Theta_{11}-\sum_{i=2}^\Ns\Theta_{ii})=\Ii\sum_{i=2}^\Ns\sum_{j=2}^\Ns (\xi_i\cdot\xi_j)^2.
\]
This is behind the condition on $a(\xi_2,\dots,\xi_\Ns)$ in \eqref{3D66a}. Using in addition the first equation \eqref{3D22a}, one can reformulate \eqref{3D66a} according to:
\begin{align} 
&a(\xi_2,\dots,\xi_\Ns)+\xi_1^\ua a_\ua(\xi_2,\dots,\xi_\Ns)\sim 0\ 
\Leftrightarrow\ \notag\\
& a(\xi_2,\dots,\xi_\Ns)=\fb(\Theta_{11}-\sum_{i=2}^\Ns\Theta_{ii})d'(\xi_2,\dots,\xi_\Ns)\ \wedge\ 
\notag\\
&\xi_1^\ua a_\ua(\xi_2,\dots,\xi_\Ns)=\fb \vartheta_1^\ua d'_\ua(\xi_2,\dots,\xi_\Ns)
\label{3D66b}
\end{align}
for some polynomials $d'(\xi_2,\dots,\xi_\Ns)$ and $d'_\ua(\xi_2,\dots,\xi_\Ns)$ in the components of $\xi_2$,\dots,$\xi_\Ns$.

\section{Conclusion}

We have computed the primitive elements of the supersymmetry algebra cohomology for supersymmetry algebras \eqref{i2-1} in $\Dim=2$ and $\Dim=3$ dimensions for all signatures $(t,\Dim-t)$ and all numbers $\Ns$ of sets of Majorana type supersymmetries (depending on the particular dimension and signature, these are Majorana-Weyl, Majorana or symplectic Majorana supersymmetries).

Thereby we have introduced methods which are applicable and useful also for analogous computations in higher dimensions. These are:
\begin{itemize}
\item "dimension-climbing", i.e. using the results in a lower dimension to derive the results in a higher dimension, cf. sections \ref{2D.1} and \ref{3D.1.1};
\item "ghost-matching" for different signatures, i.e. using appropriately defined ghost variables that allow one to match transformations and results for different signatures in a particular dimension, cf. sections \ref{2D.3} and \ref{3D.1} (see also section 5.3 of \pap);
\item "covariantization" of results, i.e. rewriting the results obtained in a particular spinor representation in an \so{t,\Dim-t}-covariant way so that they become valid for any other equivalent spinor representation, cf. sections \ref{2D.2} and \ref{3D.2}.
\end{itemize}

Furthermore the results exhibit features that will be met also in higher dimensions and are typical for the supersymmetry algebra cohomology. These are:
\begin{itemize}
\item the dependence of the primitive elements on the translation ghosts via \so{t,\Dim-t}-covariant ghost polynomials as in equations \eqref{3D20}, cf. lemmas \ref{lem3D1a} and \ref{lem3D2a};
\item a decrease of the maximal \cdeg\ (= degree in the translation ghosts) of primitive elements for increasing $\Ns$ (in $\Dim=3$ dimensions there are primitive elements with \cdeg s\ zero and one in the cases $\Ns=1$ and $\Ns=2$ but in the cases $\Ns>2$ all primitive elements have \cdeg\ zero, cf. lemmas \ref{lem3D1}, \ref{lem3D2} and \ref{lem3D3}).
\end{itemize}

As a final remark we add that the results for $\Dim=2$ apply analogously also to the case with an alternative choice of the charge conjugation matrix $\CC$ ($\sigma_1$ in place of $\sigma_2$ in the particular spinor representations given in equations \eqref{2D1}, \eqref{2D20} and \eqref{2D20a}): in lemma \ref{lem2D3} one just has to substitute $\xi^{+\ua}_{1_+} \xi^{-\ub}_{1_-}\CC^{-1}_{\ua\ub}$ for $\xi^{+\ua}_{1_+} \xi^{-\ub}_{1_-}(\GAM\CC^{-1})_{\ua\ub}$ and in lemma \ref{lem2D5} $\xi^{\ua}_{1} \xi^{\ub}_{1}\CC^{-1}_{\ua\ub}$ for $\xi^{\ua}_{1} \xi^{\ub}_{1}(\GAM\CC^{-1})_{\ua\ub}$ (cf. section 5.5 of \pap).


\begin{thebibliography}{99}
\addcontentsline{toc}{section}{References}

\bibitem{Brandt:2009xv}
  F.~Brandt,
  ``Supersymmetry algebra cohomology I: Definition and general structure,''
  \href{http://arxiv.org/abs/0911.2118}{arXiv:0911.2118} [hep-th].

\end{thebibliography}
\end{document}